\documentstyle[epsf,twoside]{article}

\catcode`\@=11
\long\def\@makefntext#1{
\protect\noindent \hbox to 3.2pt {\hskip-.9pt  
$^{{\eightrm\@thefnmark}}$\hfil}#1\hfill}		

\def\@makefnmark{\hbox to 0pt{$^{\@thefnmark}$\hss}}	
	
\def\ps@myheadings{\let\@mkboth\@gobbletwo
\def\@oddhead{\hbox{}
\rightmark\hfil\eightrm\thepage}   
\def\@oddfoot{}\def\@evenhead{\eightrm\thepage\hfil
\leftmark\hbox{}}\def\@evenfoot{}
\def\sectionmark##1{}\def\subsectionmark##1{}}

\oddsidemargin=\evensidemargin
\newcounter{sectionc}\newcounter{subsectionc}\newcounter{subsubsectionc}
\renewcommand{\section}[1] {\vspace{12pt}\addtocounter{sectionc}{1} 
\setcounter{subsectionc}{0}\setcounter{subsubsectionc}{0}\noindent 
	{\tenbf\thesectionc. #1}\par\vspace{5pt}}
\renewcommand{\subsection}[1] {\vspace{12pt}\addtocounter{subsectionc}{1} 
	\setcounter{subsubsectionc}{0}\noindent 
	{\bf\thesectionc.\thesubsectionc. {\kern1pt \bfit #1}}\par\vspace{5pt}}
\renewcommand{\subsubsection}[1] {\vspace{12pt}\addtocounter{subsubsectionc}{1}
	\noindent{\tenrm\thesectionc.\thesubsectionc.\thesubsubsectionc.
	{\kern1pt \tenit #1}}\par\vspace{5pt}}
\newcommand{\nonumsection}[1] {\vspace{12pt}\noindent{\tenbf #1}
	\par\vspace{5pt}}

\newcounter{appendixc}
\newcounter{subappendixc}[appendixc]
\newcounter{subsubappendixc}[subappendixc]
\renewcommand{\thesubappendixc}{\Alph{appendixc}.\arabic{subappendixc}}
\renewcommand{\thesubsubappendixc}
	{\Alph{appendixc}.\arabic{subappendixc}.\arabic{subsubappendixc}}

\renewcommand{\appendix}[1] {\vspace{12pt}
        \refstepcounter{appendixc}
        \setcounter{figure}{0}
        \setcounter{table}{0}
        \setcounter{lemma}{0}
        \setcounter{theorem}{0}
        \setcounter{corollary}{0}
        \setcounter{definition}{0}
        \setcounter{equation}{0}
        \renewcommand{\thefigure}{\Alph{appendixc}.\arabic{figure}}
        \renewcommand{\thetable}{\Alph{appendixc}.\arabic{table}}
        \renewcommand{\theappendixc}{\Alph{appendixc}}
        \renewcommand{\thelemma}{\Alph{appendixc}.\arabic{lemma}}
        \renewcommand{\thetheorem}{\Alph{appendixc}.\arabic{theorem}}
        \renewcommand{\thedefinition}{\Alph{appendixc}.\arabic{definition}}
        \renewcommand{\thecorollary}{\Alph{appendixc}.\arabic{corollary}}
        \renewcommand{\theequation}{\Alph{appendixc}.\arabic{equation}}
        \noindent{\tenbf Appendix \theappendixc #1}\par\vspace{5pt}}
\newcommand{\subappendix}[1] {\vspace{12pt}
        \refstepcounter{subappendixc}
        \noindent{\bf Appendix \thesubappendixc. {\kern1pt \bfit #1}}
	\par\vspace{5pt}}
\newcommand{\subsubappendix}[1] {\vspace{12pt}
        \refstepcounter{subsubappendixc}
        \noindent{\rm Appendix \thesubsubappendixc. {\kern1pt \tenit #1}}
	\par\vspace{5pt}}

\newcommand{\textlineskip}{\baselineskip=13pt}
\newcommand{\smalllineskip}{\baselineskip=10pt}

\def\eightcirc{
\begin{picture}(0,0)
\put(4.4,1.8){\circle{6.5}}
\end{picture}}
\def\eightcopyright{\eightcirc\kern2.7pt\hbox{\eightrm c}} 

\newcommand{\copyrightheading}[1]
	{\vspace*{-2.5cm}\smalllineskip{\flushleft
	{\footnotesize International Journal of Modern Physics B, #1}\\
	{\footnotesize $\eightcopyright$\, World Scientific Publishing
	 Company}\\
	 }}

\newcommand{\publisher}[2]{{\begin{center}\footnotesize\smalllineskip 
	Received #1\\
	Revised #2
	\end{center}
	}}

\def\abstracts#1#2#3{{
	\centering{\begin{minipage}{4.5in}\baselineskip=10pt\footnotesize
	\parindent=0pt #1\par 
	\parindent=15pt #2\par
	\parindent=15pt #3
	\end{minipage}}\par}} 

\renewenvironment{thebibliography}[1]			
	{\frenchspacing
	 \ninerm\baselineskip=11pt
	 \begin{list}{\arabic{enumi}.}
	{\usecounter{enumi}\setlength{\parsep}{0pt}
	 \setlength{\leftmargin 12.7pt}{\rightmargin 0pt} 
	 \setlength{\itemsep}{0pt} \settowidth
	{\labelwidth}{#1.}\sloppy}}{\end{list}}

\newcounter{itemlistc}
\newcounter{romanlistc}
\newcounter{alphlistc}
\newcounter{arabiclistc}

\newcommand{\fcaption}[1]{
        \refstepcounter{figure}
        \setbox\@tempboxa = \hbox{\footnotesize Fig.~\thefigure. #1}
        \ifdim \wd\@tempboxa > 5in
           {\begin{center}
        \parbox{5in}{\footnotesize\smalllineskip Fig.~\thefigure. #1}
            \end{center}}
        \else
             {\begin{center}
             {\footnotesize Fig.~\thefigure. #1}
              \end{center}}
        \fi}

\newcommand{\tcaption}[1]{
        \refstepcounter{table}
        \setbox\@tempboxa = \hbox{\footnotesize Table~\thetable. #1}
        \ifdim \wd\@tempboxa > 5in
           {\begin{center}
        \parbox{5in}{\footnotesize\smalllineskip Table~\thetable. #1}
            \end{center}}
        \else
             {\begin{center}
             {\footnotesize Table~\thetable. #1}
              \end{center}}
        \fi}

\def\@citex[#1]#2{\if@filesw\immediate\write\@auxout
	{\string\citation{#2}}\fi
\def\@citea{}\@cite{\@for\@citeb:=#2\do
	{\@citea\def\@citea{,}\@ifundefined
	{b@\@citeb}{{\bf ?}\@warning
	{Citation `\@citeb' on page \thepage \space undefined}}
	{\csname b@\@citeb\endcsname}}}{#1}}

\newif\if@cghi
\def\cite{\@cghitrue\@ifnextchar [{\@tempswatrue
	\@citex}{\@tempswafalse\@citex[]}}
\def\citelow{\@cghifalse\@ifnextchar [{\@tempswatrue
	\@citex}{\@tempswafalse\@citex[]}}
\def\@cite#1#2{{$\null^{#1}$\if@tempswa\typeout
	{IJCGA warning: optional citation argument 
	ignored: `#2'} \fi}}

\def\pmb#1{\setbox0=\hbox{#1}
	\kern-.025em\copy0\kern-\wd0
	\kern.05em\copy0\kern-\wd0
	\kern-.025em\raise.0433em\box0}

\def\fnt#1#2{\footnotetext{\kern-.3em
	{$^{\mbox{\scriptsize #1}}$}{#2}}}

\def\fpage#1{\begingroup
\voffset=.3in
\thispagestyle{empty}\begin{table}[b]\centerline{\footnotesize #1}
	\end{table}\endgroup}

\def\runninghead#1#2{\pagestyle{myheadings}
\markboth{{\protect\footnotesize\it{\quad #1}}\hfill}
{\hfill{\protect\footnotesize\it{#2\quad}}}}
\font\tenrm=cmr10
\font\tenit=cmti10 
\font\tenbf=cmbx10
\font\bfit=cmbxti10 at 10pt
\font\ninerm=cmr9
\font\nineit=cmti9
\font\ninebf=cmbx9
\font\eightrm=cmr8

\def\qed{\hbox{${\vcenter{\vbox{			
   \hrule height 0.4pt\hbox{\vrule width 0.4pt height 6pt
   \kern5pt\vrule width 0.4pt}\hrule height 0.4pt}}}$}}

\def\bsc{{\sc a\kern-6.4pt\sc a\kern-6.4pt\sc a}}	
\def\bflatex{\bf L\kern-.30em\raise.3ex\hbox{\bsc}\kern-.14em 
T\kern-.1667em\lower.7ex\hbox{E}\kern-.125em X} 

\begin{document}

\runninghead{Photoemission from Ordered Stripe Phases} 
{Photoemission from Ordered Stripe Phases}

\normalsize\textlineskip
\thispagestyle{empty}
\setcounter{page}{1}

\copyrightheading{}			

\vspace*{0.88truein}

\fpage{1}
\centerline{\bf 
PHOTOEMISSION FROM ORDERED STRIPE PHASES}

\vspace*{0.37truein}
\centerline{\footnotesize R.S. MARKIEWICZ and C. KUSKO}
\vspace*{0.015truein}
\centerline{\footnotesize\it Physics Department and Barnett Institute,} 
\baselineskip=10pt
\centerline{\footnotesize\it Northeastern University, Boston MA 02115, USA}
\vspace*{10pt}
\vspace*{0.225truein}
\publisher{(received date)}{(revised date)}

\vspace*{0.21truein}
\abstracts{A phase separation model for stripes has found good agreement with 
photoemission 
experiments and with other studies which suggest a termination of the striped
phase in the slightly overdoped regime.  Here the model is extended in a number 
of respects.  In particular, a discussion of the nature of the charged stripes
is presented, suggesting how density waves, superconductivity, and strong
correlations can compete with the quantum size effects inherent in narrow 
stripes.  The anomalous doping dependence of the chemical potential is 
explained.}{}{}



\vspace*{1pt}\textlineskip	
\section{Photoemission from Stripe Arrays}   
\vspace*{-0.5pt}
\noindent
Generically, any phase separation model of stripes has three characteristic
features: (i) {\it termination of the stripe phase} at some finite doping, $x_0
$; (ii) a {\it crossover} at a lower doping, $x_{cr}\sim x_0/2$ from 
magnetic-dominated ($x<x_{cr}$) to charge-dominated ($x>x_{cr}$) stripe arrays; 
(iii) some {\it interaction} on the charged stripes which stabilizes the 
particular doping $x_0$.  Recent evidence suggests that the stripes and 
pseudogap terminate at the same doping, while superconductivity 
persists\cite{Tal1}.  A consistent picture yields $x_0\simeq 0.25$, so the 
crossover can be identified with the 1/8 anomaly, where both charged and 
magnetic stripes have their minimal width (2 Cu atoms).  Then, at optimal doping
in, e.g., YBa$_2$Cu$_3$O$_{7-\delta}$ (YBCO), $x_{opt}=16/19\times 0.25=0.21
$\cite{Talx}, while the width of the charged stripes $N$ satisfies $N/(N+2)=16
/19$, or $N=32/3\sim 10$ Cu wide.

\begin{figure}\begin{center}
\leavevmode
   \epsfxsize=0.50\textwidth\epsfbox{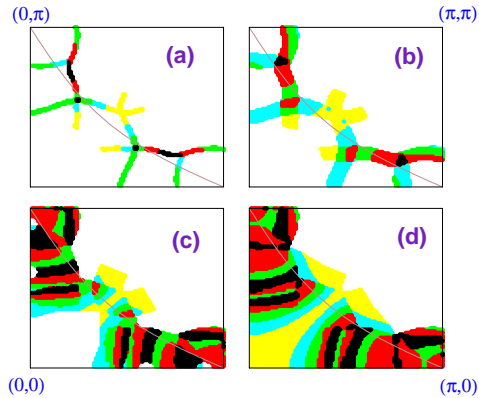}
\vskip0.5cm 
\caption{Constant energy cuts of PE dispersion for a 1/8 doping stripe
array, within (a) 30 (b) 100, (c) 200, or (d) 500 meV of the Fermi level.  The
calculation includes a matrix element, $M=|c_x-c_y|$, which suppresses intensity
along the zone diagonal.  Lines = Fermi surface of bulk (or very wide) charged 
stripes.  Relative intensity increases with darker shading.}
\label{fig:1}
\end{center}\end{figure}

Hence, models of isolated quasi-one-dimensional charged stripes are likely to be
valid only in the far underdoped regime.  To study the doping dependence of
stripes, and particularly of the wider stripes present near optimal doping, we
developed a model of ordered stripe arrays (tight-binding calculations including
Coulomb charging effects) and applied it to the study of photoemission
(PE)\cite{OSP}.  (Earlier studies of the effects of stripes on PE\cite{SEK,Sei} 
were limited to a single doping.)  We found (1) the PE consists of separate
components for the magnetic stripes (upper and lower Hubbard bands) and charged
stripes (filling in midgap states with doping); (2) the charged stripe PE 
consists of several subbands associated with quantum size effects (QSE) on the 
finite width stripes.  The two PE components are clearly resolved in La$_{2-x}
$Sr$_x$CuO$_4$ (LSCO)\cite{Ino} and constitute the peak (charged stripes) and 
hump (magnetic stripes) features in superconducting Ba$_2$Sr$_2$CaCu$_2$O$_8$ 
(BSCCO).  The main difference between the two materials is that the lower 
Hubbard band in BSCCO is considerably closer to the Fermi level, presumably an
effect of stronger stripe fluctuations.  
The doping dependence of the QSE is consistent with 
that of the pseudogap, while the intensity of the peak feature is 
found\cite{ZXTex,DEW} to scale with doping $x$, maximizing at the point where 
the stripe phase terminates, confirming that the peak is a property of the 
charged stripes.  A map of the intensity distribution near the Fermi level,
Fig.~\ref{fig:1} is in qualitatively good agreement with experiment\cite{ZZX}.

\begin{figure}\begin{center}
\leavevmode
   \epsfxsize=0.45\textwidth\epsfbox{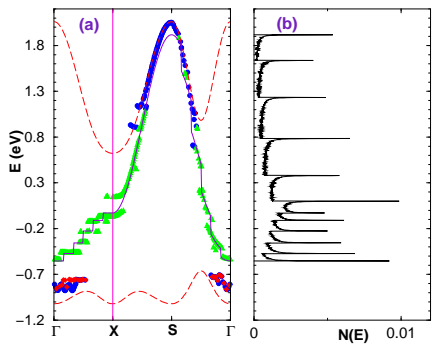}
\vskip0.5cm 
\caption{(a) Dispersion of a stripe array with charged stripes 6 Cu wide 
(magnetic stripes 2 Cu wide). Data from Fig. 7d of Ref.\protect\cite{OSP}; 
triangles (diamonds) = predominantly from charged (magnetic) stripes, while
circles = mixed origin; dashed line = Mott bands of magnetic stripes; solid
line = single (charged) stripe model, with $k_x$ approximated by nearest 
quantized value.  (b) Density of states for a single stripe 6 Cu wide.}
\label{fig:2}
\end{center}\end{figure}

\vspace*{1pt}\textlineskip	
\section{Nature of Charged Stripes}   
\vspace*{-0.5pt}
\noindent
For the stripe phase to exist, the doping $x_0$ must be particularly {\it 
stable}.  This can arise via an electronic {\it instability}, which opens up a
gap over much of the Fermi surface, making the electronic phase nearly
incompressible.  This `Stability from Instability' is a fairly general feature,
underlying, e.g., Hume-Rothery alloys\cite{HumR}.  [This is a modification of an
argument due to Anderson\cite{And}.]
Here, we explore a number of candidates for the predominant electronic
instability.  To simplify the calculation, we note that the PE from the charged
stripes in an array is well modelled by the PE from an isolated charged ladder
of the same width, Fig.~\ref{fig:2}.  [Note from the dos that the Van Hove 
singularity remains well defined on a stripe.]  Hence, we need only study 
instabilities on ladders.  

We explored the competition between a charge-density wave (CDW) and d-wave
superconductivity on a ladder in a weak coupling calculation similar to 
Ref.~\cite{MKK,BFal}.  For wide stripes, the bulk results are recovered.
As the stripe width decreases, quantum confinement rapidly eliminates the CDW
gap, by $N\sim 8$.  Superconductivity is less affected, but is still suppressed
by $N=2$.  [If the doping on the stripe were not fixed, there would be a large
superconducting gap when the Fermi level coincides with the one-dimensional VHS 
of a stripe subband.]  

\begin{figure}\begin{center}
\leavevmode
   \epsfxsize=0.40\textwidth\epsfbox{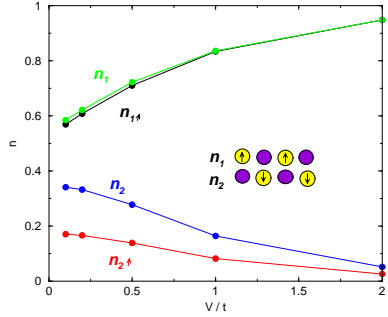}
\vskip0.5cm
\caption{Linear antiferromagnetic (LAF) array with CDW, showing spin and doping
distribution on different sites as a function of interaction strength $V$, 
assuming $U=6t$, $t'=-0.276t$.
Inset shows arrangement of atoms.}
\label{fig:20b}
\end{center}\end{figure}

Strong correlations can be included by incorporating some kind of spin ordering
on the stripes.  At the mean-field level, we have found a low-energy, phase 
separated solution to the Hubbard model\cite{MarKI} which closely resembles a
White-Scalapino (WS) stripe\cite{WhiSc}.   In a one-band Hubbard model with 
mean-field magnetization $m_q$, the quasiparticle dispersion is 
\begin{equation}
E_{\pm}={\epsilon_k+\epsilon_{k+q}\over 2}\pm 
\sqrt{({\epsilon_k-\epsilon_{k+q}\over 2})^2+U^2m_q^2},
\label{eq:4}
\end{equation}
with 
\begin{equation}
\epsilon_k=-2t(c_x+c_y)-4t'c_xc_y,
\label{eq:1}
\end{equation}
and $c_i=\cos{k_ia}$.  For the cuprates, we expect\cite{OSP} $t\simeq 325meV$, 
$U/t\simeq 6$ and $t'/t\simeq -0.276$.  For $q$ = $\vec Q\equiv (\pi ,\pi )$, 
this is the dispersion we assumed for the antiferromagnetic (AFM) stripes. 
A linear antiferromagnetic (LAF) phase arises when $q$ = $(\pi ,0)$; in general
its properties closely resemble those of the WS stripes.  For instance, 2-Cu
wide LAF stripes act as antiphase boundaries for AFM stripes, a finite $t'$
destabilizes the LAF phase, and the hole doping on an LAF stripe is close to
that on a WS stripe\cite{MarKI}.  The Fermi surfaces for AFM-LAF stripe arrays
are even closer to experiment\cite{ZZX} than those of Fig.~\ref{fig:1}.

\begin{figure}\begin{center}
\leavevmode
   \epsfxsize=0.40\textwidth\epsfbox{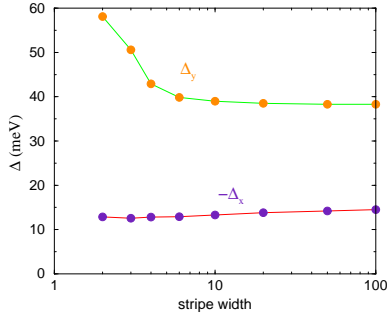}
\vskip0.5cm    
\caption{Linear antiferromagnetic (LAF) array with `d-wave' superconductivity, 
showing magnitude of gap along ($y$) or across ($x$) the stripes, as a function
of stripe width.}
\label{fig:20c}
\end{center}\end{figure}

Ordered phases are much more stable on LAF ladders.  For instance, near $x=0.5$ 
there is a CDW phase stabilized by near neighbor Coulomb repulsion $V$, which is
highly stable, essentially independent of ladder width.  For this strongly 
correlated CDW, the hole density varies from $~0$ to 1, not 2, 
Fig.~\ref{fig:20b}.  An attractive $V$ can stabilize a d-wave-like 
superconductor on a ladder, with an anisotropic gap Fig.~\ref{fig:20c} which 
actually increases for the narrowest stripes.  The optimum superconducting gap
corresponds to the Fermi level at the LAF saddle point.

\vspace*{1pt}\textlineskip	
\section{Chemical Potential Shifts in a Stripe Phase}   
\begin{figure}\begin{center}
\leavevmode
   \epsfxsize=0.45\textwidth\epsfbox{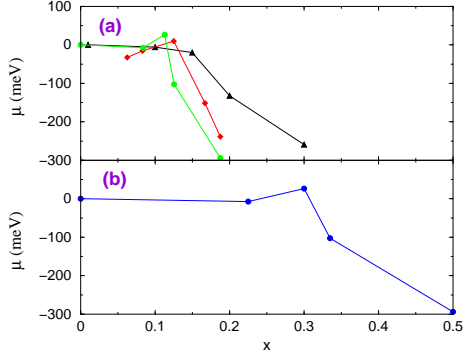}
\vskip0.5cm 
\caption{Doping dependence of chemical potential $\mu$ for (a) 
LSCO (triangles\protect\cite{In1}) and (b) LSNO\protect\cite{Sat}.  In (a), the
circles are the scaled data of LSNO, while the diamonds are calculated from
stripe band shifts associated with charging effects (see text).}
\label{fig:3}
\end{center}\end{figure}

\vspace*{-0.5pt}
\noindent
While stripes persist up to $x_0=0.25$ in LSCO, the chemical potential is 
independent of doping\cite{In1} only between half filling ($x=0$) and $x=0.125$,
Fig.~\ref{fig:3}a.  Actually, the same anomaly is found in La$_{2-x}$Sr$_x$NiO$_
4$ (LSNO)\cite{Sat}, Fig.~\ref{fig:3}b, which has long-range charge order up to
at least $x=0.5$, although $\mu (x)$ is constant only up to $x_c=0.33$.  
By rescaling the LSNO $x_c$ to that of the cuprates (circles in 
Fig.~\ref{fig:3}a), it can be seen that the dependence $\mu (x/x_c)$ is quite 
similar in the two compounds.  This also suggests an explanation: the break in 
slope for $\mu$ above $x_c$ is associated with commensurability effects.  Each
commensurate configuration has a well defined Madelung energy due to charge
inhomogeneity.  Between two commensurate configurations this charging energy
changes linearly with doping (the intermediate states are presumed to be 
mixtures of the commensurate phases), but crossing over a commensurate phase
leads to a different mixed phase, and a change in slope of the charging energy.
In a layered compound, this charging energy contributes to the chemical 
potential of the layer involved.  Thus, in the nickelates, the break is near 1/3
doping, and the 1/3 stripes are found to be stable over an extended doping 
range.  In the cuprates, a similar effect at 1/8 is very plausible, since the 
charging energy is minimized at that doping\cite{OSP}.  In fact, we can estimate
the charging energy, since this Coulomb interaction also raises the chemical 
potential of the charged stripes with respect to the magnetic stripes, as found 
in our earlier calculation\cite{OSP}.  These two charging effects should be
proportional. The diamonds in Fig.~\ref{fig:3}a plot the calculated\cite{OSP} 
band-edge shift of the charged stripes with respect to the magnetic stripes, 
showing a reasonable agreement with the CuO$_2$ plane chemical potential shift.

\nonumsection{Acknowledgements}
\noindent
These computations were carried out using the facilities
of the Advanced Scientific Computation Center at Northesatern University 
(NU-ASCC).  Their support is gratefully acknowledged.  We thank M.A.H.
Vozmediano and Z.-X. Shen for stimulating conversations.
Publication 788 of the Barnett Institute.

\nonumsection{References}

\end{document}